\documentstyle[preprint,aps,eqsecnum]{revtex}
\begin{document}

\title
{Phonon Mediated Transresistivity in a Double Layer Composite Fermion System}

\author{D.V. Khveshchenko}
\address
{NORDITA, Blegdamsvej 17, Copenhagen DK-2100, Denmark}

\maketitle

\begin{abstract}
\noindent
We consider the fractional drag in a 
double layer system of two-dimensional electrons in the half-filled lowest Landau level. At
sufficiently large inter-layer separations the drag is dominated by exchange of acoustic phonons 
and exhibits novel temperature and inter-layer distance 
dependences. At low temperatures the phonon mediated drag is 
strongly enhanced with respect to the case of zero magnetic field.
\end{abstract}
\pagebreak

In the recent years, the problem of two-dimensional electrons in a double 
quantum well has attracted a lot of, both, experimental and theoretical attention. 
Apart from other interesting properties, this system offers a new opportunity for studying
electron correlations by virtue of the phenomenon of the frictional drag. 
The latter manifests itself in the form of a finite 
transresistivity defined as a voltage induced in an open circuit (passive) layer 
while a current is flowing through the other (active) one (see \cite{JS} and references therein).

In the absence of electron tunneling,
the transresistivity which takes its origin
in momentum transfer from the active to the passive layer can only result
from inter-layer Coulomb coupling. 
Thus, being of the interaction origin, the transresistivity provides a sensitive probe for such important
characteristics of the two-dimensional electron gas as the electron polarization operator   
and the dynamically screened inter-layer coupling.

Detailed experimental studies of the frictional drag in semiconductor heterostructures
revealed features associated with inter-layer 
plasmons \cite{G}. However, it was predicted that, despite 
its dominance at small inter-layer distances 
$d$, the drag caused by the direct Coulomb coupling decays rapidly with $d$ \cite{JS}. 
Since the experimentally measured drag showed 
a weaker distance dependence at large $d$, it was suggested that some other mechanism 
of momentum transfer ought to be at work. 

As such, exchange of acoustic phonons was proposed as the most likely alternative.
Improving on the earlier results obtained in \cite{G,TVP}, the authors of  
Ref.\cite{BFHM} considered the effects of, both, deformation potential (DP)  
and piesoelectric (PE) electron-phonon couplings. This work was primarily concerned with 
the behavior at temperatures comparable to or higher than the Bloch-Gruneisen temperature $T_{BG}=2uk_F$
defined in terms of the Fermi momentum $k_F$ and the speed of sound 
$u$, where the DP coupling is known to be far more important than the PE one.
At temperatures smaller than $T_{BG}$ their relative strength reverts though.

According to the general argument \cite{JS}, the low-temperature transresistivity
is controlled by the phase-space volume available for 
thermally excited bosonic modes of an appropriate kind. Thus, measuring an exponent $n$ 
in the low-temperature 
dependence $\rho_{12}\sim T^n$ allows one to identify a mechanism responsible for the drag.
Namely, the Coulomb drag gives $n=2$ while the phonon mediated drag tends to yield  
substantially higher $n$ values \cite{BFHM}. 

Since at high enough temperatures the frictional drag of any origin is expected to
undergo a crossover to a universal linear $T$ dependence \cite{JS}, 
an observation of a maximum of $\rho_{12}(T)/T^2$ at $T\approx T_{BG}$ \cite{G} 
provides a strong evidence in favor of the phonon mechanism.  

Recent drag measurements performed in the Quantum Hall regime \cite{L} revealed new features 
which are believed to be associated with a formation of compressible metal-like electron states
at half-filling as well as other even-denominator filling fractions $\nu\sim 1/2p$ 
of the lowest Landau level. In the framework of the theory by Halperin, Lee, and Read \cite{HLR} 
these states are described in terms of spinless
fermionic quasiparticles dubbed composite fermions (CFs).
In the mean field picture of the best studied $\nu=1/2$ Quantum Hall state,
the CFs experience zero effective field and occupy all states
inside some ostensible Fermi surface of the size $k_F=(4\pi n_e)^{1/2}$ given in terms of the electron density
$n_e$. 

Among other differences from ordinary electrons 
in zero field, the compressible CF states
were predicted to feature a peculiar Coulomb drag behavior $\rho_{12}\sim T^{4/3}$   
characteristic of an anomalously slow relaxation of overdamped density fluctuations \cite{S,US1}.

Further work in this direction focused on the experimentally observed 
tendency of $\rho_{12}(T)$ to saturate at low $T$ or even increase upon decreasing
the driving current \cite{L}, the features that were attributed to a possible CF inter-layer pairing which
might result in a formation of an incompressible (intrinsically double-layer) gapful electron state 
\cite{US2}.

However, on the general grounds, one expects that neither the direct Coulomb coupling  
nor the proposed pairing mechanism remain relevant at large enough inter-layer separations.
Instead, the drag measurements performed in this regime may reveal important
details of coupling between the CFs and acoustic phonons. 

Motivated by this expectation, in the present paper we
consider the phonon mediated drag in a widely-separated double layer 
CF system.

Earlier insight into the problem of the CF-phonon coupling
has been gained from the analysis of the available experimental data on phonon-limited mobility \cite{K}, 
phonon drag contribution to thermopower \cite{T}, and energy loss rate by hot electrons due
to phonon emission \cite{C} at $\nu=1/2$.

The bare coupling between CFs and PE phonons is described by the standard electron-phonon vertex
$$
M_{\lambda}({\bf Q})=eh_{14}(A_{\lambda}/2\rho u_{\lambda}Q)^{1/2}, \\\ 
A_{l}={9q^2_zq^4\over 2Q^6}, \\\ A_{tr}={8q^4_zq^2+q^6\over 4Q^6}
\eqno(1)$$
where ${\bf Q}=({\bf q}, q_z)$ is the 3D phonon momentum, $\rho$ is the bulk density of $GaAs$, $u_{\lambda}$
is the speed of sound with polarization $\lambda =l, tr$, and $h_{14}$
is the non-zero component of the piesoelectric tensor which relates the local electrostatic
potential to the lattice displacement.

As demonstrated in Ref.\cite{KR}, the bare PE vertex (1) undergoes full dynamical screening
 governed by   
the intra-layer $V_{11}(q)=2\pi e^2/\epsilon_0q$ 
and inter-layer 
$V_{12}(q)=V_{11}(q) e^{-qd}$ Coulomb potentials,
$\epsilon_0\approx 12.9$ being the bulk dielectric constant of $GaAs$.

In the CF theory, the effects of screening are described in terms of
the intra-layer CF density $\Pi_{00}(\omega, {\bf q})$ and current $\Pi_{\perp}(\omega, {\bf q})$ 
polarization functions. However, when computing physical quantities, 
these polarizations always appear in a particular combination, which one can readily recognize as 
the irreducible electron density response function \cite{US1} 
$$
\chi(\omega,{\bf q})={\Pi_{00}\over {1-(4\pi)^2\Pi_{00}\Pi_{\perp}}}={q^2\over {q^2(dn_e/d\mu)^{-1}
-(4\pi)^2 i\omega\sigma(q)}}
\eqno(2)$$
This peculiar form of the density response which is characteristic of the compressible CF states    
is given in terms of the CF compressibility $dn_e/d\mu$ and the momentum-dependent
conductivity $\sigma(q)={1\over 2}k_F min (l, 2/q)$, where $l$ stands for the CF mean free path \cite{HLR}. 
Throughout this paper, we refer to the momenta $q$ larger (smaller) than $l^{-1}$ as ballistic 
(diffusive) regime, respectively.

Moreover, as a straightforward analysis shows,
the possibility to express such quantities of physical interest as the transresistivity solely
in terms of $\chi(\omega,{\bf q})$ also holds for the dynamically screened
CF-phonon coupling included alongside with the direct inter-layer Coulomb interaction. 

With both present, the two interactions constitute the effective inter-layer coupling 
$$
W_{12}={V_{12}+D_{12}\over {(1+\chi (V_{11}+D_{11}))^2-\chi^2(V_{12}+D_{12})^2}}
\eqno(3)$$
where the two-dimensional electron-phonon interaction function $D_{ij}(\omega,{\bf q})$ is computed with the 
use
of the bulk phonon propagator ($\omega_+=\omega+i(u/2l_{ph})sgn\omega$)  
$$
D_{\lambda}(\omega, {\bf Q})=
{2u_{\lambda}Q\over {\omega_+^2-u^2_{\lambda}Q^2}}, 
\eqno(4)$$ 
which depends on the three-dimensional momentum ${\bf Q}=({\bf q}, q_z)$
and accounts for a finite phonon mean free path $l_{ph}$ due to boundary scattering,  
and the wave function of the lowest occupied transverse electron subband $\psi(z)\sim ze^{-z/w}$
that determines the formfactor $F_i(q_z)=\int dz |\psi_i(z)|^2e^{iq_zz}$.

For $\omega$ close to $u_{\lambda}q$ the relevant values of $q_z$ turn out to be small.
Therefore one can only keep the contribution of 
transverse phonons $(A_l\ll A_{tr}\approx 1/4)$ with velocity $u\approx 3\times 10^5 cm/s$ 
which for $qw\lesssim 1$ results in the formula \cite{BFHM}:  
$$
D_{ij}(\omega,{\bf q})=\sum_{\lambda}\int {dq_z\over 2\pi}F_i(q_z)F_j(q_z)
|M_{\lambda}({\bf Q})|^2
{\cal D}_{ij}(\omega,{\bf Q}) $$
$$\approx 
-{(eh_{14})^2\over 8u^2\rho 
{\sqrt {q^2-(\omega_+/u)^2}}}(\delta_{ij}+(1-\delta_{ij})e^{-d{\sqrt {q^2-(\omega_+/u)^2}}})
\eqno(5)$$
Since perpendicular magnetic field does not affect the $z$-axis confinement,
Eq.5 is applicable at arbitrary magnetic fields.

To the (lowest nontrivial) second order in the effective inter-layer coupling (3)
the transresistivity is given by the standard expression \cite{JS,US1}:
$$
\rho_{12}={1\over 8\pi^2}{h\over e^2}{1\over Tn^2_e}\int {d{\bf q}\over (2\pi)^2}
\int^{\infty}_0 d\omega  {\large (}
{q Im \chi(\omega,{\bf q}) \over {\sinh{\omega\over 2T}}}{\large )}^2 
 |W_{12}(\omega,{\bf q})|^2
\eqno(6)$$
which can also incorporate, under some modifications, the effects of a magnetic field and/or disorder.

In order to facilitate a direct comparison with the calculation of $\rho_{12}(T)$ in zero field, 
we will refer to the following expression 
$$\epsilon(\omega, {\bf q})={dn_e\over d\mu}(\chi^{-1}+ V_{11})
=1+{dn_e\over d\mu}(V_{11}(q)+{i\omega\sigma(q)\over q^2})
\eqno(7)$$
as the effective dielectric function of a single layer CF system.

By making use of Eq.7 and putting $D_{12}\approx D_{11}=D$ we arrive at the expression  
$$
\rho_{12}={1\over 16\pi^3}{h\over e^2}{1\over Tn^2_e}\int{\sigma^2(q)d{q}\over q}
\int {\omega^2d\omega \over \sinh^2{\omega\over 2T}}
 {|V_{12}(q)+D(\omega,{\bf q})|^2\over {|(\epsilon(\omega,{\bf q})-V_{12}(q))
(\epsilon(\omega,{\bf q})+V_{12}(q)+2D(\omega,{\bf q}))|^2}}
\eqno(8)$$
where the upper limit in the integral over momentum is set by either the maximum span of the CF Fermi surface
$2k_F$ or by the inverse width of the quantum well $w^{-1}$, which, for the sake of simplicity, 
we choose to be equally restrictive ($2k_Fw\sim 1$).

The relative strength of the PE phonon coupling is controlled
by a small parameter $\eta=h_{14}^2\epsilon_0/16\pi u^2\rho\sim 10^{-3}$.
Therefore it proves to be convenient to divide
the $q$-integral onto the momenta smaller and greater than $d^{-1}$.
In the former, the small parameter $\eta$ allows one
to neglect $D$ in Eq.8 altogether and, in this way, to recover the pure case of the Coulomb drag. 
In this range of momenta, the integral over frequencies receives its main contribution 
from $\omega\sim q^2V_{11}(q)/\sigma(q)$ corresponding to the overdamped
density mode \cite{US1,S} whose (purely imaginary) dispersion is derived from the equation 
$\epsilon(\omega,{\bf q})=0$.

Depending on the values of the two dimensionless parameters $\xi=k_Fd$ and $\sigma=k_Fl$, 
the contribution of the 
momenta $q\lesssim d^{-1}$ undergoes a number of crossovers.
Provided that both $\sigma$ and $\xi$ are sufficiently large,
Eq.8 yields a variety of regimes characterized by different dependences on 
the dimensionless temperature $\tau=T/\Delta$ measured in units 
of the average Coulomb energy $\Delta=k_Fe^2/\epsilon_0\sim 100 K$:
$$
\rho^C_{12}(T)\propto
\pmatrix{(\tau\sigma/\xi)^2\ln(\tau\sigma^3/\xi),\\\ \tau<\xi/\sigma^3 \cr
(\tau/\xi)^{4/3},\\\ \xi/\sigma^3<\tau<1/\xi^2 \cr
\tau/\xi^2,\\\ 1/\xi^2<\tau} or \\\
\pmatrix{(\tau\sigma/\xi)^2\ln(\tau/\sigma\xi),\\\ \tau<1/\sigma\xi \cr
\tau\sigma/\xi^3,\\\ 1/\sigma\xi<\tau}
\eqno(9)$$
in the situations $l>d$ and $l<d$, respectively.

In the case of a long CF mean free path,
the asymptotical low-$T$ regimes $\rho^C_{12}\sim T^{4/3}$ and $T^2\ln T$ were previously
obtained under the assumptions of a ballistic and diffusive CF dynamics, respectively \cite{US1}.
According to Eq.9, upon increasing the inter-layer separation beyond the CF mean free path
the former regime ceases to be accessible. 

Since neither Eq.9 nor its zero field counterpart $\rho^{C,0}_{12}\sim \tau^2\xi^{-4}$ \cite{JS} 
depend on the mass of the charge carriers, 
 the only reason behind the observed three orders of magnitude enhancement 
of the $\nu=1/2$ drag at $d\approx 300 A$ \cite{L} is because is was measured at low enough $T$. 
 
As follows from Eq.9, when treated in the second order perturbation theory,   
the pure Coulomb drag at $\nu=1/2$ can only be responsible for a temperature dependence $\rho_{12}(T)\sim T^n$ 
with $n<2$, including the regime of the logarithmic enhancement. 

Since the inter-layer Coulomb correlations decay rapidly with $d$, and the perturbative result (9) becomes 
more
and more accurate,
one might conclude that the Coulomb drag alone could not explain a maximum 
in $\rho_{11}(T)/T^2$, should one be observed.

Nonetheless, the recent experiment on the double layer $\nu=1/2$ system 
with $d\approx 5000 A$ did reveal such a maximum  occurring at roughly
the same $T\sim 2 K$ as in the zero field case \cite{Z}.
By analogy with the situation in zero field, this observation may 
call for an alternate mechanism of inter-layer momentum transfer.

In our unified approach that led to Eq.8, an additional contribution due to exchange of PE phonons (both real
and virtual ones corresponding to the contributions proportional to 
 $ImD$ and $ReD$, respectively \cite{TVP,BFHM})
comes from the momenta $q\gtrsim d^{-1}$, at which one 
can instead neglect $V_{12}$ in the integrand.

First, we consider the case of a short phonon mean free path due to boundary scattering
(the exact conditions under which this approximation holds are to be specified later).

The phonon mediated contribution to the transresistivity can be cast in the form
$$
\rho^{ph}_{12}\approx {1\over 16\pi^3}{h\over e^2}{\eta^2\over Tn^2_e}\int^{2k_F}_{1/d} {\sigma^2(q)dq\over q}
\int^{\infty}_0 {\omega^2d\omega\over \sinh^2 {\omega\over 2T}}
{e^{-d{\sqrt {(q^2-(\omega_+/u)^2)}}}\over |\epsilon(\omega,{\bf q})|^4
{\sqrt {(q^2-(\omega/u)^2)^2+(q/l_{ph})^2}}}
\eqno(10)$$
By introducing the parameter $\delta=u\epsilon_0/e^2\approx 10^{-2}$ we present
the results of the integrations in the form 
$$
\rho^{ph}_{12}(T)/\eta^2\ln(l_{ph}/d) \propto
\pmatrix
{\tau^4\sigma^2\delta^{-2}(1+\sigma^2\delta^2)^{-2},
\\\ \delta\xi^{-1}< \tau <\delta\sigma^{-1} \cr
\tau^6\delta^{-8}, \\\ max (\delta\xi^{-1}, \delta\sigma^{-1})< \tau < \delta^2 \cr
\tau^2,\\\ max (\delta\xi^{-1}, \delta\sigma^{-1}, \delta^2)< \tau < \delta \cr
\tau\delta, \\\ \delta < \tau }
\eqno(11)$$
All of the above regimes can only be accessible
 if there exist sizable intervals of temperature within the bounds
imposed by the conditions  $\xi^{-1}\ll \sigma^{-1}\ll \delta\ll 1$. 
With all the regimes present, the exponent $n$ in the approximate power-law temperature dependence
of $\rho^{ph}_{12}(T)$ does not remain constant and changes non-monotonously as $T$ gets smaller.

The $\ln (l_{ph}/d)$ dependence of the transresistivity on the inter-layer separation holds 
as long as $d\lesssim l_{ph}$,
while at higher $T$ Eq.11 yields $\rho^{ph}_{12}\sim \exp(-d/l_{ph})$. 

A different variety of regimes occurs in the case of a long phonon mean free path
where, in order to arrive at a finite result, one has to keep $D$ in the denominator of Eq.8
that now reads as
$$
\rho^{ph}_{12}\approx {1\over 16\pi^3}{h\over e^2}{\eta^2\over Tn^2_e}\int_{1/d}^{2k_F} {\sigma^2(q)dq\over q}
\int^{\infty}_0 {\omega^2d\omega\over \sinh^2{\omega\over 2T}}
 {e^{-d{\sqrt {q^2-(\omega/u)^2}}}
 \over {|\epsilon(\omega,{\bf q})(\epsilon(\omega,{\bf q}){\sqrt {q^2-(\omega/u)^2}}-2\eta)|^2}}
\eqno(12)$$
With $D$ included, the integrand in (12) develops a new weakly damped pole at $\omega \approx uq{\sqrt 
{1-\eta^2}}$,  
which corresponds to a collective electron-phonon mode that forms in a double layer system (cf.\cite{BFHM}).
When integrating over frequencies, we restrict ourselves to the vicinity of this pole.

The coupled electron-phonon mode can only occur provided
that neither the real nor the imaginary part
of the denominator in (12) is affected by the intrinsic phonon lifetime at all $q\gtrsim d^{-1}$.
The corresponding criteria can be expressed as  
$l_{ph}\gg \eta^{-2}max( d, (k_F\delta)^{-1}, d\sigma^{-1}\delta^{-1})$.

Expanding the denominator 
near the pole and
carrying out the integrations  we arrive at the following regimes:
$$
\rho^{ph}_{12}(T)/\eta^2 \propto
\pmatrix
{\tau^4\sigma\delta^{-3}(1+\sigma^2\delta^2)^{-1}, \\\ \delta\xi^{-1}< \tau <\delta\sigma^{-1} \cr
\tau^5\delta^{-6}, \\\ max (\delta\xi^{-1}, \delta\sigma^{-1})< \tau <\delta^2 \cr
\tau^3\delta^{-2},\\\ max (\delta\xi^{-1}, \delta\sigma^{-1}, \delta^2)< \tau < \delta \cr
\tau, \\\ \delta < \tau} 
\eqno(13)$$
Unlike the Coulomb drag (9) that originates from small momenta $q\lesssim d^{-1}$, 
the integral (12) receives only 
a negligible correction from the region where the dielectric function $\epsilon(\omega,{\bf q})$ approaches 
zero.

According to Eq.13, the contribution of the electron-phonon mode remains nearly independent of
$d$ as long as the condition $d{\sqrt {q^2-(\omega/u)^2}}\approx qd\eta\lesssim 1$
is satisfied for all momenta less than $2k_F$, which implies $ \xi\eta<1$.

Similar to the case of ordinary electrons in zero field
 Eq.12 includes, apart from the Lorenzian-like contribution associated with the collective mode
 that Eq.13 does account for, a substantial non-Lorenzian tail due to processes involving virtual phonons, 
 which gives rise to a non-trivial overall dependence on $d$ (cf. Ref.\cite{BFHM}). For this reason
one should not expect the estimates (11) and (13) 
obtained under the assumptions of, respectively, short  and long $l_{ph}$ 
to match smoothly at $l_{ph}\sim d\eta^{-2}$.

In order to simplify a comparison between our results and the zero field analysis 
 \cite{BFHM}
we make the following choice of the parameters: $\xi\sim \sigma\sim\delta^{-1}$,
 which greatly reduces the number of the relevant regimes.  
Incidentally, the above choice is not that far
off the experimental situation of Ref.\cite{Z}: $d\approx 5000A$ and $l\sim 1\mu m$
which corresponds to a single layer resistivity $\rho_{11}\sim 2000\Omega$.

At these parameter values, in the range of temperatures
 $0.01 \lesssim T/T_{BG} \lesssim 1$ one can only expect to observe the asymptotical behavior 
$\rho^{ph}_{12}(T)\sim \eta^2(T/T_{BG})^2\delta^2\ln(l_{ph}/d)$ 
or $\eta^2(T/T_{BG})^3\delta$  for the 
phonon mean free path being shorter or longer than its crossover value estimated as $d\eta^{-2}$,
which is well above a typical one $l_{ph}\sim 0.1mm$ \cite{T}. 

In either case, the exponent is substantially
reduced as compared to the case of ordinary electrons in zero field: 
$\rho^{ph,0}_{12}\sim \eta^2(T/T^0_{BG})^6\delta^2\ln(l_{ph}/d)$ and $\eta^2(T/T^0_{BG})^5\delta$,
respectively.

Thus, compared to its zero field counterpart, the phonon mediated drag in the half-filled
Landau level is strongly enhanced at all temperatures smaller than $T_{BG}$
(in Ref.\cite{Z} the measured ratio of the transresistivities
 was $\rho^0_{12}/\rho_{12}\sim 10^{-3}$ at $T\approx 1K$),
while they become comparable at $T\sim T_{BG}$.

For lower values of the exponent $n$ in the power-law temperature dependence
of the transresistivity the maximum of the ratio $\rho_{12}(T)/T^2$ shifts towards lower $T$.
 When contrasted against its position in zero field, this downward shift could, at least partly, compensate 
for 
 another, 
upward, shift due to the extra factor $T_{BG}/T^0_{BG}={\sqrt 2}$ resulting
from the ratio between the CF and the ordinary electron's Fermi momenta at a given density \cite{HLR}.
Interestingly enough, in Ref.\cite{Z} the maximum was found to occur in both cases at roughly the same 
position,
which was, however, lower than the estimated value of $T^0_{BG}\approx 5K$.

Compared to the direct Coulomb drag in the CF system (9),
the relative strength of the
phonon mediated drag is weaker than in the case of ordinary electrons in zero field.
Indeed, on the basis of Eq.13 one would conclude
that even at $T\sim T_{BG}$ the two competing contributions to the transresistivity can only become
comparable at $\xi\sim \eta^{-1}$ which 
corresponds to inter-layer separations in excess of $10\mu m$, while 
the use of Eq.11 would result in even higher values of $d$.

Conversely, at zero field the phonon-mediated drag outpowers the direct Coulomb one already at $\xi\gtrsim 
\delta^{1/4}
\eta^{-1/2}$ which only requires separations $d\gtrsim 10^3A$.

Thus, it does not seem to be obvious that 
 under the experimental conditions of Ref.\cite{Z} the phonon mediated drag
indeed dominates over the direct Coulomb one 
in the CF regime, although it indeed appears to be the main 
contribution at zero field and $T\sim T^0_{BG}$. 

It is worthwhile mentioning that, by itself, an 
observation of a peak in $\rho_{12}(T)/T^2$ in a magnetic field 
may not yet provide an unambiguous evidence in favor of the phonon mechanism. In 
Ref.\cite{BFHJ} such a peak was predicted to occur
in strong enough magnetic fields with no reference to phonons or   
the peculiar CF behavior in the lowest Landau level.
Seemingly corroborating, is the observation \cite{Z}  
of the peak's evolution with magnetic field at fixed $k_F$ in a whole range of magnetic fields
(see also \cite{R}).

Finally, a comparison between the data from Refs.\cite{L} and \cite{Z} shows that upon increasing $d$ by a 
factor
of $\sim 20$ both $\rho^0_{12}(1K)$ and $\rho_{12}(1K)$ drop by the factors of $\sim 50$ and $\sim 250$,
respectively.  

In the case of $\rho^0_{12}$, this change is markedly less pronounced than the $\sim \xi^{-4}$ behavior
of the direct Coulomb drag \cite{JS}, which could be naturally explained
if a crossover to the weakly $d$-dependent phonon mediated regime occurred already at $d\sim 10^3A$.

On the contrary, the change in $\rho_{12}$ is somewhat more consistent with
the roughly $\sim \xi^{-2}$ behavior of the CF Coulomb
drag than with the nearly $d$-independent phonon contribution that has taken over at some intermediate $d$.
Thus, despite some qualitative agreement with the experimental findings of Ref.\cite{Z},   
our results call for further experimental studies of the frictional drag in the CF regime, which have to
be performed at even larger inter-layer separations $d$ in order to unequivocally point at the phonon-related 
origin
of the phenomenon. 

To summarize, we carried out a comprehensive analysis of the frictional drag
in the double layer compressible CF state
of electrons in the lowest Landau level. We considered the effect of the inter-layer electron-phonon coupling 
and 
contrasted it against the drag caused by the direct Coulomb interaction.
We found that in the CF system
the low temperature phonon mediated drag is strongly enhanced compared to 
the case of ordinary electrons in zero field. 
However, in order to single out the phonon drag
in the CF system, the transresistivity measurements have to be done at 
larger inter-layer distances than those which are generally sufficient
in the zero field case.

\end{document}